\documentclass[english,prl,twocolumn,superscriptaddress]{revtex4-1}
\usepackage[T1]{fontenc}
\usepackage[latin9]{inputenc}
\usepackage{color}
\usepackage{amssymb}
\usepackage{graphicx}

\makeatletter
\@ifundefined{textcolor}{}
{%
 \definecolor{BLACK}{gray}{0}
 \definecolor{WHITE}{gray}{1}
 \definecolor{RED}{rgb}{1,0,0}
 \definecolor{GREEN}{rgb}{0,1,0}
 \definecolor{BLUE}{rgb}{0,0,1}
 \definecolor{CYAN}{cmyk}{1,0,0,0}
 \definecolor{MAGENTA}{cmyk}{0,1,0,0}
 \definecolor{YELLOW}{cmyk}{0,0,1,0}
}

\makeatother

\usepackage{babel}
\begin{document}

\title{Interpreting the macroscopic pointer by analysing the Einstein-Podolsky-Rosen
steering of an entangled macroscopic superposition state}

\author{{\normalsize{}M. D. Reid }}

\affiliation{Centre for Quantum Optical Science, Swinburne University of Technology,
Melbourne, Australia}
\begin{abstract}
We examine Einstein-Podolsky-Rosen's (EPR) steering nonlocality
for two realisable Schrodinger cat-type states where a meso/ macroscopic
system (called the ``cat''-system) is entangled with a spin-$1/2$
system.  For large cat-systems, we show that a local hidden state
model is near-satisfied, meaning that the cat-system can be consistent
with being in a mixture of ``dead'' and ``alive'' states despite
that it is entangled with the spin system. We justify that a rigorous
signature of the Schrodinger cat-type paradox is the EPR-steering
of the cat-system and provide two experimental signatures. This leads
to a hybrid quantum/ classical interpretation of the macroscopic pointer
of a measurement device and suggests many Schrodinger cat-type
paradoxes can be explained by microscopic nonlocality.
\end{abstract}
\maketitle
The original arguments of Einstein-Podolsky-Rosen (EPR) and Bell dealt
with small symmetrical systems: two particles or two spins \cite{epr-1,bell2}.
The arguments are based on EPR's notion of local realism (LR) $-$
put simply, that there can be no ``spooky action-at-a-distance''
\cite{einsteinspooky} on one system as a result of measurements made
on the other. In revealing inconsistencies between the predictions
of quantum mechanics and the premise of local realism (LR), these
arguments have had profound implications for physics \cite{brunner}.
Schrodinger recognised that the consequences of such paradoxes would
be significant for larger systems \cite{Schrodinger-1}. He analysed
a quantum gedanken experiment whereby a macroscopic system $C$ (likened
to a cat and that we refer to as the ``cat-system'') becomes entangled
with a microscopic spin $1/2$ system $S$, the final state being
the superposition 
\begin{equation}
|\psi\rangle=\frac{1}{\sqrt{2}}\Bigl(|A\rangle_{C}|\uparrow\rangle_{S}+|D\rangle_{C}|\downarrow\rangle_{S}\Bigr)\label{eq:cat}
\end{equation}
Here, $|A\rangle_{C}$ and $|D\rangle_{C}$ represent macroscopically
distinguishable states in which the ``cat'' is ``dead'' (if given
by $|D\rangle_{C}$) or ``alive'' (if given by $|A\rangle_{C}$).
The $|\uparrow\rangle$, $|\downarrow\rangle$ are the eigenstates
of the Pauli spin $\sigma_{Z}$. The spin and ``cat'' systems can
in principle become spatially separated. 

While Schrodinger pointed out the natural interpretation of this state
$-$ that the ``cat'' cannot be viewed as either ``dead'' or ``alive''
until measured $-$ he did not construct an EPR-type experiment that
would \emph{demonstrate} such failure of reality for a practical realisation
of (1). While such signatures have since been developed (for example
\cite{cat bell,NOON bell bana,LG,svet-1,multiprticle}), experimental
work has mainly focused on providing evidence (such as a fidelity
or entanglement measure) for the state (1) within quantum theory \cite{cats,ions,NOON,svetexp,catcqed,legexp}
and the signatures do not directly examine the reality of the cat-system
$C$ itself, as distinct from that of the spin-system $S$. Yet,
understanding the precise nature of the failure of classical realism
for the state (1) is of topical interest: Many proposals have been
put forward so that the paradoxical situation in which a ``cat''
is apparently \emph{'both alive and dead'} can be better understood
\cite{peres,bohm,bellbook}.

The objective of this Letter is to probe the asymmetrical nature of
the entanglement of the superposition state (1) by way of an EPR
paradox (called an EPR steering paradox in this more general situation
\cite{steer-1-1}). The important feature of our analysis is that
EPR's local realism is defined \emph{asymmetrically}, so one may consider
either ``spooky'' action on the cat-system\emph{ }$C$ by measurements
on the spin-system $S$, or vice versa \cite{epr-1}. This opens
up the possibility that nonlocality (which is the negation of LR)
can manifest asymmetrically between the two systems \cite{asymm,optoHe=000026Reid}.
That EPR-steering can be detectable one-way but not the other has
recently been confirmed experimentally for qubit systems \cite{asymmexp}.

In this Letter, we utilise this feature to gain an understanding of
the discrepancy between the quantum and classical descriptions (which
we call the degree of ``\emph{quantumness}'') for \emph{each} of
the sub-systems (the ``cat'' $C$ and the spin $S$) of the state
(1). This information is not given by the observation of entanglement
alone. It would be expected that this discrepancy can be different
for the two sub-systems. We find this is indeed the case, but are
further able to show that the quantumness of the cat-system can approach
zero, to the point where surprisingly the cat-system can be described
as ``dead'' or ``alive'', despite that the two systems remain
entangled. In this limit, the cat-system acts as a \emph{classical}
\emph{measuring device} for the microscopic system, which maintains
its quantumness. 

This motivates the question of how to determine when the cat-system
itself is paradoxical, along the lines suggested by Schrodinger. Such
a bound is not set at the realisation of entanglement, but (we show)
is set by the realisation of an EPR steering \emph{of the cat-system}.
This type of EPR steering manifests as a falsification of certain
hidden  states \emph{for the cat-system}, that are implied by the
premise of LR. These hidden states (or ``elements of reality'',
as EPR called them) \emph{predetermine} results for measurements
on the ``cat''. In this paper, we calculate details of
such elements of reality for two realisations of (1) one involving
coherent states and the other Greenberger-Horne-Zeilinger (GHZ) spin
states. By revealing contradictions, we thus arrive at measurable
signatures for the EPR steering of the cat-system for two experimentally
realisable mesoscopic superposition states. We confirm that as the
cat-system becomes larger, the hidden states become indistinguishable
from classical states (in which the cat-system is ``alive'' or ``dead'').
In this limit, the EPR steering of the cat-system is eventually lost,
but we verify that EPR steering of the spin by the cat-system can
be maintained, so that measurements on the cat-system can certify
the quantumness of the spin. 

The regime where the cat-system is large is particularly interesting,
since in this limit the cat-system models a pointer of a measuring
device for $\sigma_{Z}$ of the spin. Measurement paradoxes have
been raised in the literature, because the interpretation of the ``cat''
being both ``alive'' and ``dead'' is then that the ``\emph{pointer}''
is at two (macroscopically separated) positions of a dial at once.
While decoherence mechanisms preclude such a result, it is interesting
nonetheless to understand the entangled state \emph{without decoherence}
and how the collapse of the pointer into a state of ``one position
\emph{or} the other'' occurs. Our results indicate that the ``\emph{quantum
pointer}'' regime is very nearly a regime of one-way steering (or
entanglement) where the cat-system is fully classical. We discuss
how this suggests a hybrid quantum-classical picture of the \emph{pointer},
that immediately after interaction with the microscopic system, the
pointer is located at (near) one of the two macroscopically distinct
positions, but with an indeterminacy related to nonlocal effects 
bounded in size by the uncertainty relation.

\textbf{\emph{Coherent cat-states: }}Consider the following well-known
prototype for the superposition state (\ref{eq:cat}) \cite{cats}:
\begin{equation}
|\psi_{coh}\rangle=\frac{1}{\sqrt{2}}\Bigl(e^{-i\pi/4}|\alpha\rangle|\uparrow\rangle_{Z}+e^{i\pi/4}|-\alpha\rangle|\downarrow\rangle_{Z}\Bigr)\label{eq:classic cat haroche}
\end{equation}
Here $|\alpha\rangle$ is a coherent state for a quantum harmonic
oscillator system that we refer to as the ``cat''-system $C$. The
$|\uparrow\rangle_{Z}$, $|\downarrow\rangle_{Z}$ are eigenstates
of $\sigma_{Z}$ for the spin system $S$. We take $\alpha$ to be
real and (ideally) large. Observers Alice and Bob can make measurements
on the spin and cat-systems respectively. We consider that the two
systems have become spatially separated after the interaction that
created the entanglement. If Alice measures $\sigma_{Z}$ and the
result is $1$, then the state of system $C$ is $|\alpha\rangle$.
Similarly, if the result is $-1$, the state is $|-\alpha\rangle$.
Thus, Alice can \emph{predict} the statistics for Bob's measurements,
conditional on her outcome. Suppose Bob makes a measurement of either
the position or momentum quadrature defined (in a rotating frame)
by $X=\frac{1}{\sqrt{2}}(a^{\dagger}+a)$ and \textcolor{black}{$P=\frac{i}{\sqrt{2}}(a-a^{\dagger})$.
Here $a^{\dagger}$, $a$ are the creation, destruction operators
for system $C$. }$ $If Alice's outcome is $\pm1$, the conditional
probability distribution $P(x)$ for the outcome $x$ of Bob's measurement
$X$ in each case is the Gaussian hill 
\begin{equation}
P_{\pm}(x)=\frac{1}{\sqrt{\pi}}\exp{\{-(x\mp\sqrt{2}\alpha)^{2}\}}\label{eq:gaushill}
\end{equation}
centred at $\pm\sqrt{2}\alpha$ respectively and with variance $(\Delta x)^{2}=1/2$
as for a coherent state. The $\pm$ hills are distinguished as ``alive''
and ``dead'' for large $\alpha$.

EPR postulated that the measurement by Alice makes no difference to
the system of the other observer. Bell's expression of LR is that
the joint probability $P_{CS}(x,y)$ for outcomes $x$ and $y$ of
measurements made at $C$ and $S$ respectively can be described by
a Local Hidden Variable (LHV) model such that \cite{brunner,bell2,bellbook}
\begin{equation}
P_{CS}(x,y)=\int_{\lambda}d\lambda\rho(\lambda)P_{C}(x|\theta,\lambda)P_{S}(y|\phi,\lambda)\label{eq:lhv}
\end{equation}
Here $\lambda$ symbolises a set $\{\lambda\}$ of hidden variables
that have a distribution $\rho(\lambda)$; $\phi$ and $\theta$ are
the measurement choices for $S$ and $C$ respectively. The locality
assumption is that $P_{C}(x|\theta,\lambda)$ is independent of Alice's
measurement choice $\phi$ (for spin) and the outcome $y$ at location
$S$; similarly $P_{S}(y|\phi,\lambda)$ is independent of Bob's choice
$\theta$ and result $x$ at $C$. We note there is an \emph{asymmetry}
in the locality assumption for the EPR experiment, because the measurements
$x$ and $\theta$ by Bob are spacelike separated from those of Alice
but are in the future \cite{supp}. We call this premise LR $ $$S\rightarrow C$.
Of special interest to us is where an extra constraint is put on the
$P_{C}(x|\theta,\lambda)$ that they be consistent with the statistics
arising from a local quantum state i.e. that there exists a quantum
density operator $\rho_{C,\lambda}$ that predict the probabilities
$P_{C}(x|\theta,\lambda)$. Such probabilites are denoted with a subscript
$q$, and the model becomes the Local Hidden State (LHS) model of
Ref. \cite{steer-1-1}
\begin{equation}
P_{CS}(x,y)=\int_{\lambda}d\lambda\rho(\lambda)P_{C}(x|\theta,\lambda)_{q}P_{S}(y|\phi,\lambda)\label{eq:lhs}
\end{equation}
the falsification of which is \emph{certification of EPR-steering
of the cat-system $C$}. 

EPR noticed that the assumption of LR and a strong statistical correlation
between two systems $S$ and $C$  place restrictions on the hidden
variables $\lambda$ and the predictions $P_{C}(x|\theta,\lambda)$
given in (\ref{eq:lhv}). We find that the local cat-system must be
consistent with being in a mixture of hidden states that predetermine
the cat-system to be either ``dead'' or ``alive''. This is expressed
as the following result, proved in the Supplemental Materials \cite{supp}. 

\textbf{Result (1a)}\textbf{\emph{:}} Given LR (as the LHV model
(4)), the local cat-system $C$ is consistent with being \emph{either}
in a (hidden-variable) state with the distribution for $X$ given
by $P_{+}(x)$ (``alive''), \emph{or }in a state with statistics
given by $P_{-}(x)$ (``dead''). This implies that the hidden variable
set $\{\lambda\}$ includes a variable $\lambda_{Z}$, which defines
the two predetermined states by $\lambda_{Z}=+1$ or $-1$ respectively.
EPR used the term ``elements of reality'' to describe the predetermination.

To show EPR-steering, we consider that Alice measures $\sigma_{X}$
\cite{eric }. Alice is able to predict the probability distribution
for Bob's measurement $P$ on the system $C$, conditional on her
outcome $\pm1$. The conditional distribution $P(p)$ for $P$ is
\begin{eqnarray}
P_{\pm}(p) & = & \frac{1}{\sqrt{\pi}}\exp(-p^{2})\bigl(1\pm\sin(2\sqrt{2}p\alpha)\bigr)\label{eq:pintX}
\end{eqnarray}
 The distribution exhibits interference fringes and has a variance
$(\Delta p)^{2}=\frac{1}{2}-2\alpha^{2}e^{-4\alpha^{4}}$ reduced
\emph{below} that of the coherent state, for which $(\Delta p)^{2}=\frac{1}{2}$.
Result (1a) leads to the conclusion the cat-system $C$ is in one
or other of two states, that correspond to the distributions $P_{+}(p)$
and $P_{-}(p)$ respectively. We denote these hidden states by the
variable $\lambda_{X}$, which assumes the value $+1$ or $-1$ in
each case. For consistency with the LHV model (\ref{eq:lhv}), we
show ((\textbf{\emph{Result 1b}}) in the Supplemental Materials) that
the local cat-system $C$ would \emph{simultaneously} be described
by both variables: $\lambda_{Z}$ and $\lambda_{X}$. We represent
such an element of reality state by the ordered pair $(\lambda_{Z},\lambda_{X})$.

Now we note the inconsistency that gives an EPR steering paradox.
There are\emph{ four} element of reality states of the cat, as depicted
in Figure 1: each ($\lambda_{Z},\lambda_{X})$ has predictions for
$X$ and $P$ given by $P_{\lambda_{Z}}(x)$ and $P_{\lambda_{X}}(p)$
respectively. We see that for each of these states 
\begin{equation}
\Delta X\Delta P=\frac{1}{2}\bigl(1-4\alpha^{2}e^{-4\alpha^{4}}\bigr){}^{1/2}<\frac{1}{2}\label{eq:epruncert}
\end{equation}
which contradicts the Heisenberg uncertainty relation $\Delta X\Delta P\geq\frac{1}{2}$.
Thus, the element of reality states cannot be \emph{quantum} states:
The inequality (7) is the EPR steering inequality $\Delta_{inf}X\Delta_{inf}P<1/2$
where $(\Delta_{inf}X)^{2}=\sum_{\sigma_{Z}}P(\sigma_{Z})(\Delta(X|\sigma_{Z}))^{2}$
and $(\Delta_{inf}P)^{2}=\sum_{\sigma_{X}}P(\sigma_{X})(\Delta(P|\sigma_{X}))^{2}$
are the average inference variances for $X$ and $P$. Here, $P(\sigma_{Z})$
is the probability of outcome $\sigma_{Z}$ for $\sigma_{Z}$ and
$(\Delta(X|\sigma_{Z}))^{2}$ is the variance of the conditional distribution
$P(X|\sigma_{Z})$. This inequality signifies the failure of all LHS
models (\ref{eq:lhs}) and hence an an \emph{EPR steering of the cat-system}
\cite{eric ,eprr-2,steer-1-1}. 

In other words, the inequality (7) negates that the local cat-system
$C$ is in any mixture of any ``dead'' or ``alive'' \emph{(local)
quantum} states as consistent with the LHV model (\ref{eq:lhv}).
This is proved for all $\alpha$. However, as $\alpha\rightarrow\infty$,
the falsification of the LHS model (\ref{eq:lhs}) (evident by the
fringe pattern) becomes unverifiable. This is shown in Figure 2, where
for $\alpha\sim100$, the LHS model cannot be falsified visually given
the finite resolution of the graphics. 

The main point of this paper is that where the LHS model (\ref{eq:lhs})
is not falsifiable, there can be \emph{no demonstration of the loss
of classical reality of the cat-system itself}. This is because the
expression (\ref{eq:lhs}) describes the cat-system being in a classical
mixture of the local hidden quantum states $\rho_{C,\lambda}$ consistent
the hidden variable $\lambda_{Z}$ and therefore being in quantum
states either ``dead'' or ``alive''. It is known that the LHS
model (\ref{eq:lhs}) can hold, despite that the two systems are entangled
\cite{steer-1-1}. Entanglement is certified by negating the quantum
separable model where the predictions $P_{S}(y|\phi,\lambda)$ are
also constrained to be consistent with a quantum density operator.
In short, entanglement can be confirmed based on a strong nonclassicality
of the spin system $S$, regardless of the quantumness of the cat-system,
and is a less rigorous measure of the cat-paradox.

\begin{figure}
\includegraphics[width=0.25\columnwidth]{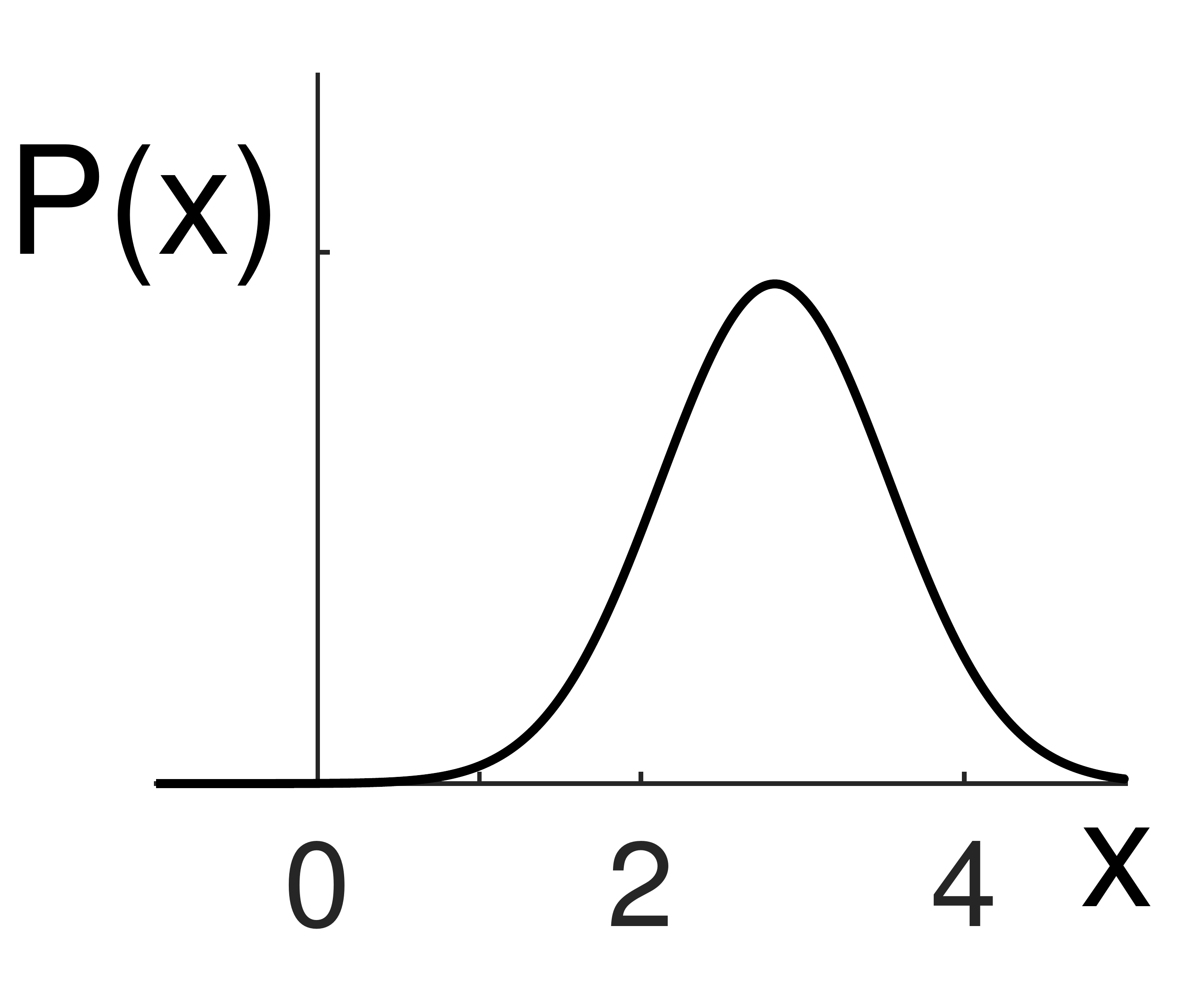}\includegraphics[width=0.25\columnwidth]{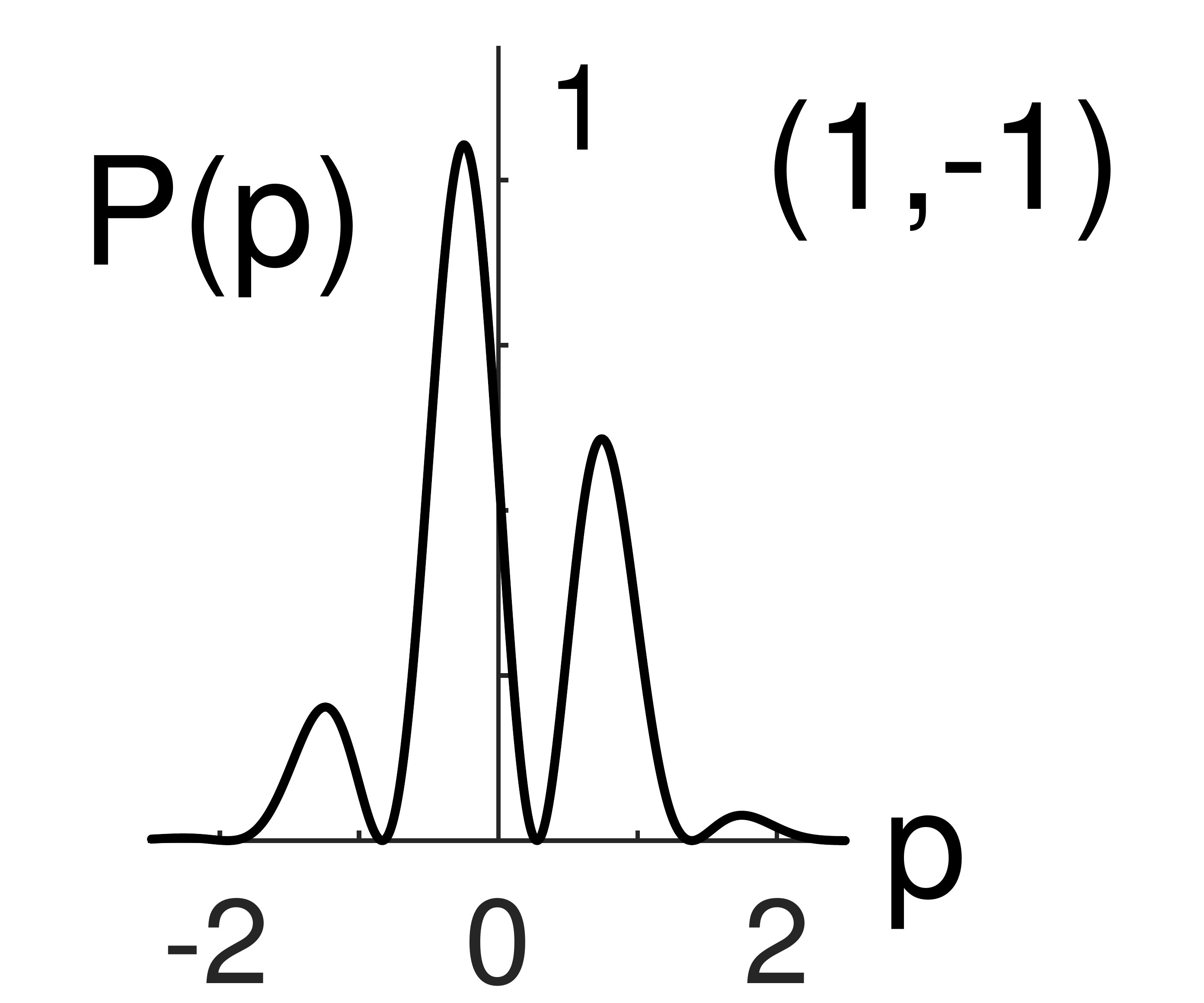}\includegraphics[width=0.25\columnwidth]{fig1a}\includegraphics[width=0.27\columnwidth]{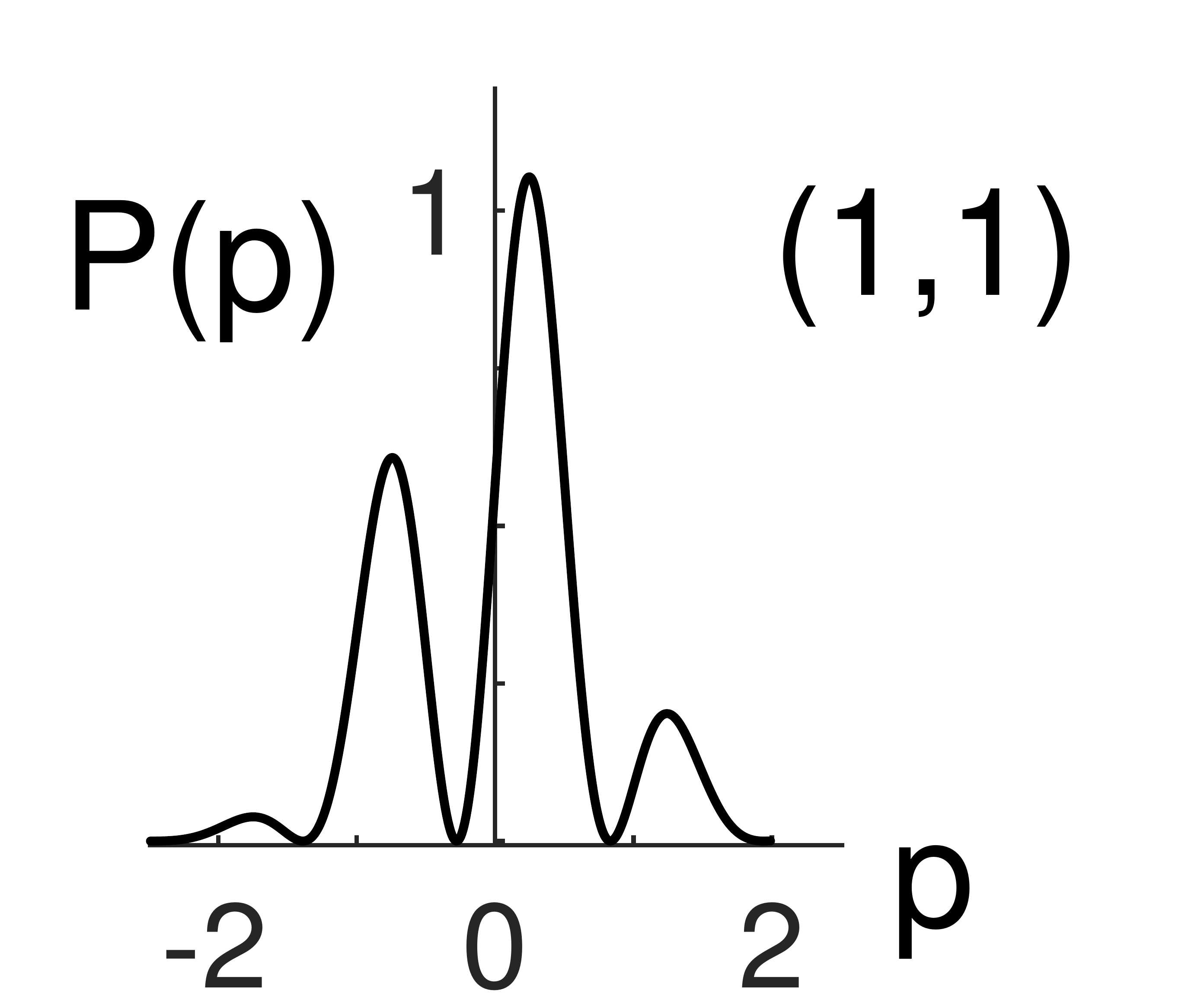}

\smallskip{}

\includegraphics[width=0.25\columnwidth]{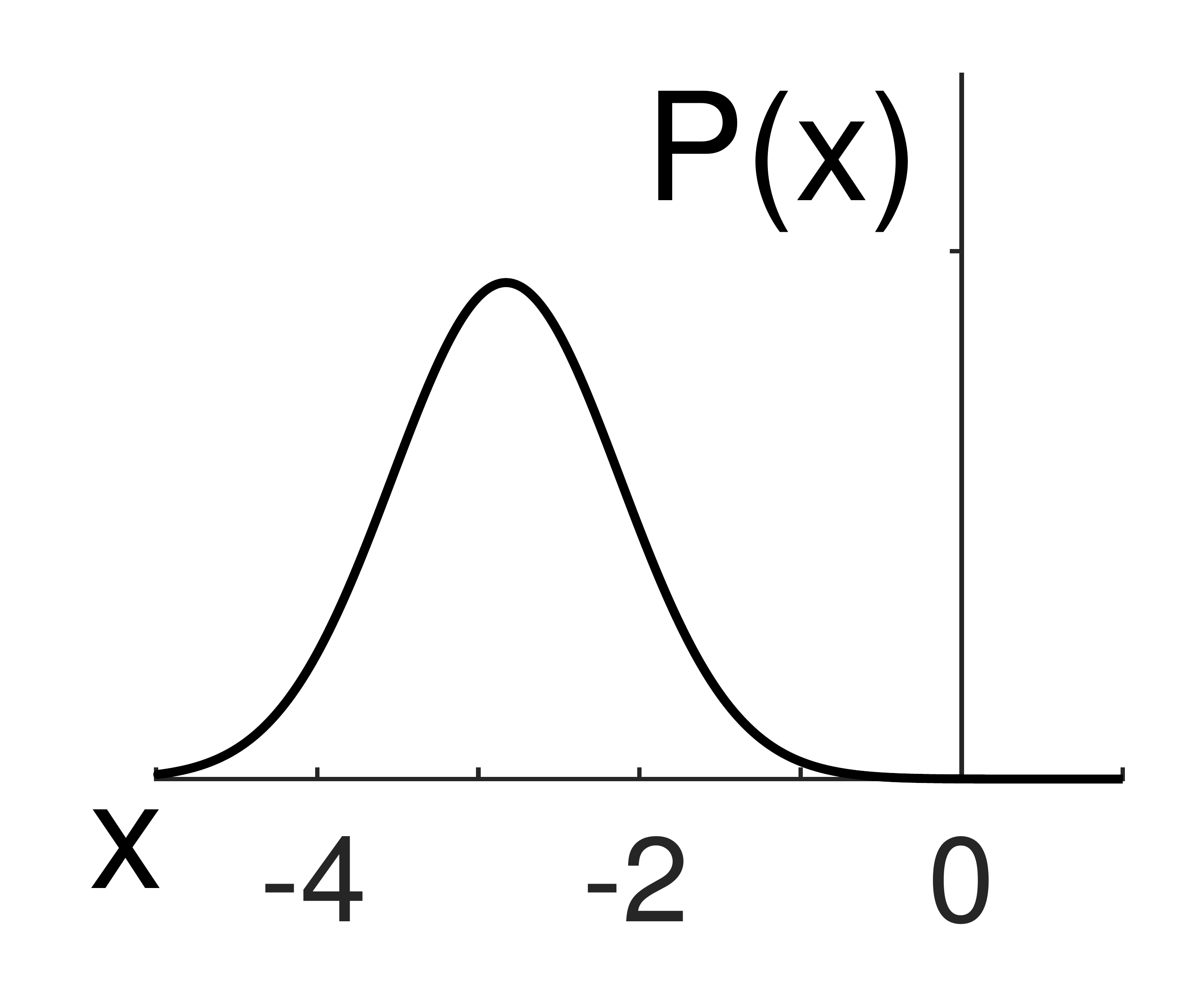}\includegraphics[width=0.25\columnwidth]{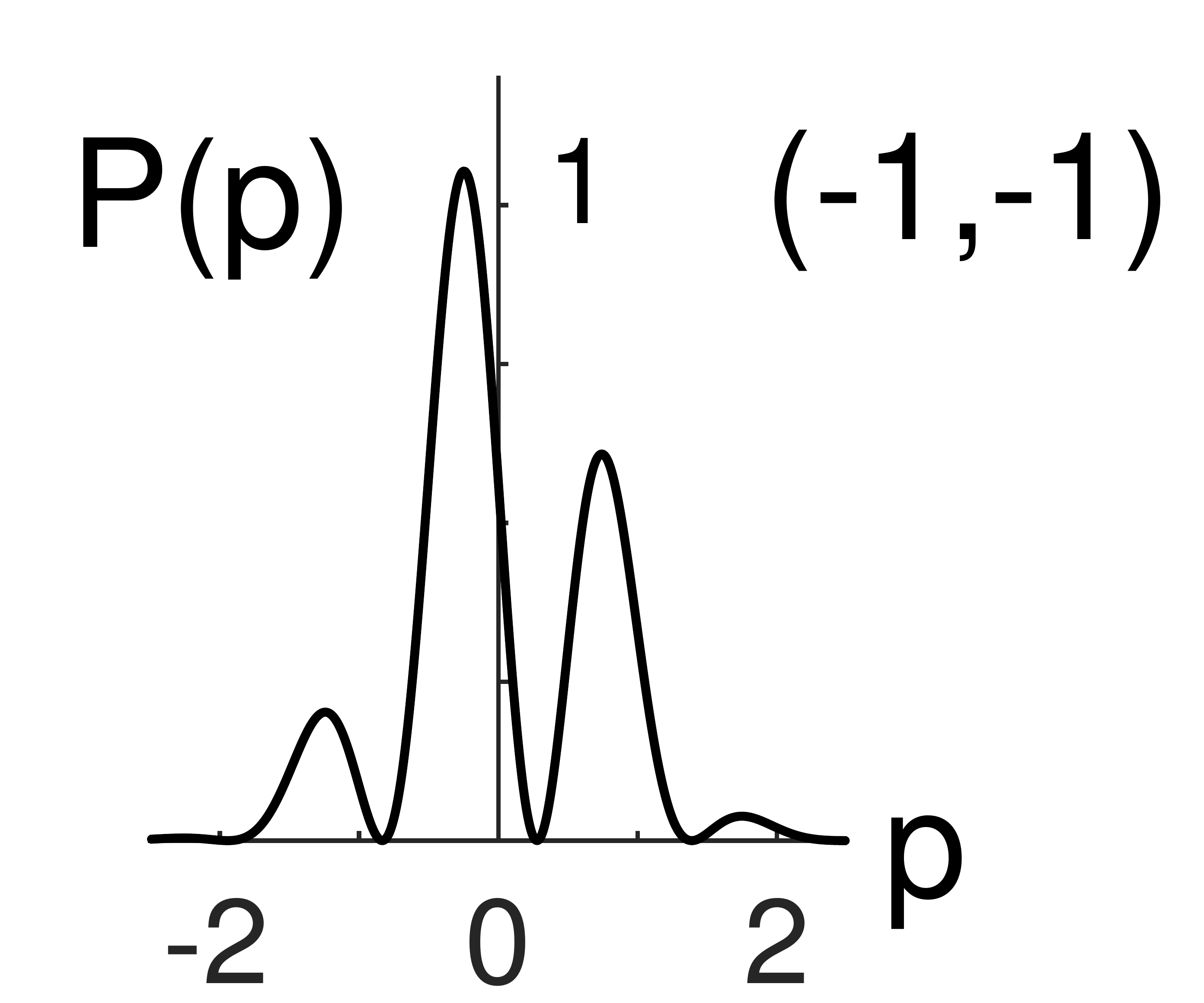}\includegraphics[width=0.25\columnwidth]{fig1d}\includegraphics[width=0.25\columnwidth]{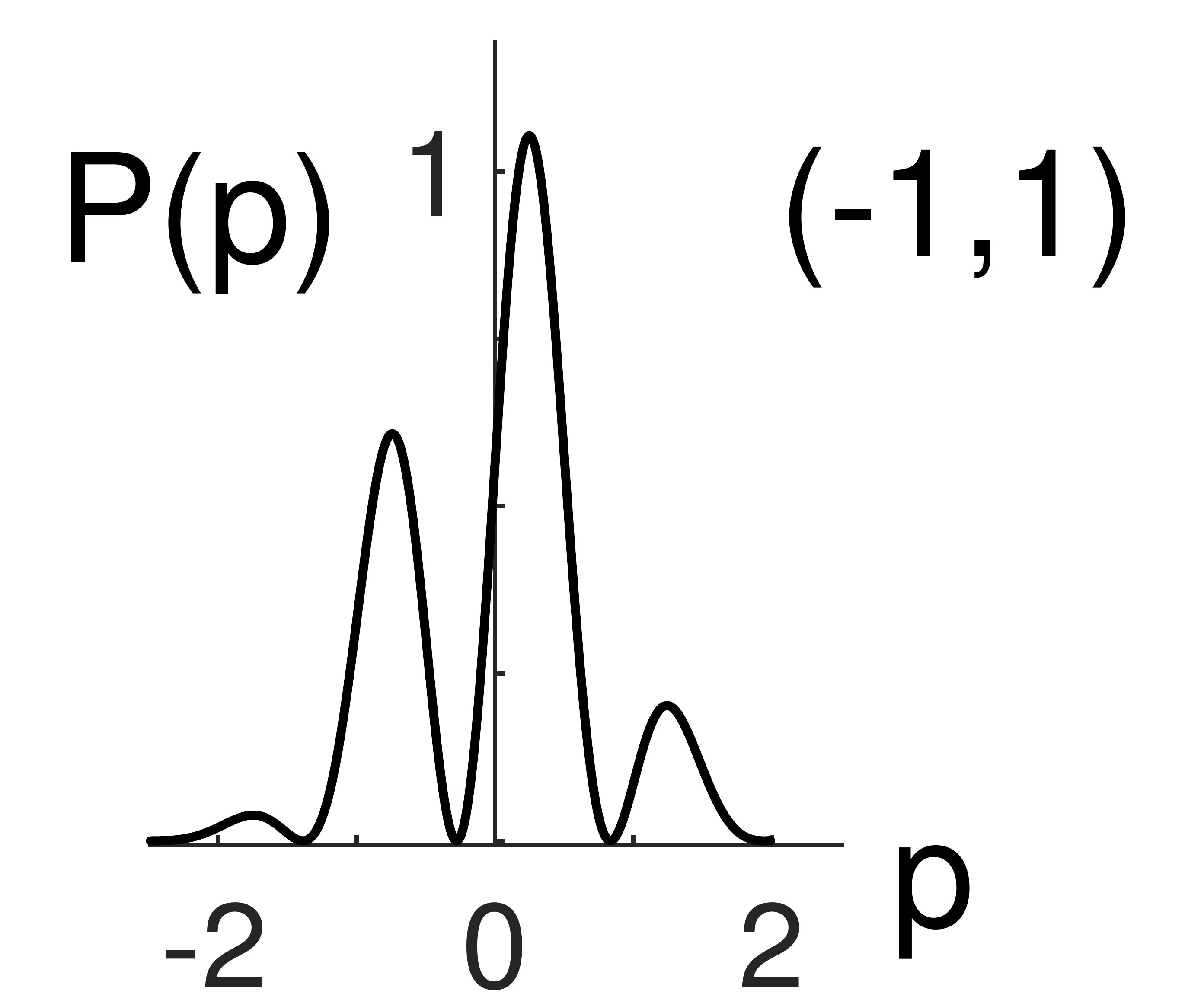}

\protect\caption{Predictions $P(x)$ and $P(p)$ for ``element of reality'' states
$(\lambda_{Z},\lambda_{X})$ of the coherent cat-state (2) with $\alpha=2$.
Top: The ``alive'' cat-system with $(1,-1)$ or $(1,1)$. Lower:
The ``dead'' cat-system with $(-1,-1)$ or $(-1,1)$.}
\end{figure}

From (2) we see that as $\alpha\rightarrow\infty$, measurement of
$X$ is also a measurement of spin $\sigma_{Z}$. The EPR steering
of the spin-system $S$ can be realised by number measurements on
system C that distinguish between adjacent odd and even values. Thus,
operational regimes of genuine one-way steering in which a fully classical
(``dead'' or ``alive'') cat-system can detect of the ``quantumness''
of the spin (but not vice versa) are viable. The full details are
given in the Supplemental Materials \cite{supp}. 

\textbf{\emph{GHZ states:}} Similar results are achieved for a second
realisation of the state (1). The GHZ state $|\psi_{GHZ}\rangle=\frac{1}{\sqrt{2}}\bigl(|\uparrow\rangle^{\otimes N}-|\downarrow\rangle^{\otimes N}\bigr)$
is formed from $N$ spin-$1/2$ particles \cite{ions,svetexp,merminghz}:
Here $|\uparrow\rangle^{\otimes N}=\prod_{k=1}^{N}|\uparrow\rangle^{(k)}$
and $|\uparrow\rangle^{(k)}$ is the spin eigenstate for $\sigma_{Z}^{(k)}$,
the $\sigma_{Z}$ observable for the $k$-th particle. If we separate
the $N$-th spin from the remaining $N-1$ spins, the GHZ state is
a microscopic spin $S$ entangled with a larger system $C$ similar
to (1): 
\begin{equation}
|\psi_{GHZ}\rangle=\frac{1}{\sqrt{2}}\Bigl(|\uparrow\rangle_{C}^{\otimes N-1}|\uparrow\rangle^{(N)}-|\downarrow\rangle_{C}^{\otimes N-1}|\downarrow\rangle^{(N)}\Bigr)\label{eq:ghz-1-1}
\end{equation}
Alice makes a measurement on the single spin, while Bob measures the
``cat''-system $C$ of $N-1$ particles. We define the collective
spin for the system $C$ as $\sigma_{Z}^{B}=\sum_{k=1}^{N-1}\sigma_{Z}^{(k)}$.
The measurement of $\sigma_{Z}^{(N)}$ by Alice will reduce the
cat-system into the ``alive'' state $|\uparrow\rangle^{\otimes N-1}$
if her result is $+1$, or to the ``dead'' state $|\downarrow\rangle^{\otimes N-1}$
if her result is $-1$. Assuming LR and Result 1, the cat-system
is deduced to be ``alive'' \emph{or }``dead'' i.e. always in one
\emph{or} the other of two element of reality states that correspond
to the outcomes $\pm(N-1)/2$ for $\sigma_{Z}^{B}$ respectively.
We denote these respective states by a hidden variable $\lambda_{Z}$
with values $\pm1$. To realise EPR steering, we consider that
Alice measures $\sigma_{X}^{(N)}$. We choose $N=3,7,..$ A measurement
of $\sigma_{X}^{(N)}$ gives the result $1$ or $-1$  which predicts
precise outcomes for Bob's $Pr_{Y}^{B}=\Pi_{k=1}^{N-1}\sigma_{Y}^{(k)}$.
Assuming LR, Result 1 implies system $C$ to be in one of the element
of reality states specified by a hidden variable $\lambda_{X}$, where
the value $\lambda_{X}=\pm1$ corresponds to outcomes for $Pr_{Y}^{B}$
 being $\pm1$. Suppose Alice measures $\sigma_{Y}^{(N)}$. Assuming
LR, the system $C$ is also in an element of reality state denoted
by a third hidden variable $\lambda_{Y}$ where the value $\lambda_{Y}=\pm1$
corresponds to outcomes for all products $Pr_{Y}^{B}(J)=\sigma_{X}^{(J)}\Pi_{k=1,k\neq J}^{N-1}\sigma_{Y}^{(k)}$,
$J=1,..N-1$ being $\pm1$. Therefore, Result 1b implies the cat-system
$C$ to be in one of the simultaneous element of reality states $(\lambda_{Z},\lambda_{X},\lambda_{Y})$
in which the outcomes for $\sigma_{Z}^{B}$, $Pr_{Y}^{B}$, $Pr_{Y}^{B}(J)$
are each predetermined with no uncertainty. Yet, observables $\sigma_{Z}^{B}$,
$Pr_{Y}^{B}$, $\sum_{J=1}^{N-1}Pr_{Y}^{B}(J)$ satisfy the Heisenberg
uncertainty relation 
\begin{equation}
\Delta(\sigma_{Z}^{B})\Delta(Pr_{Y}^{B})\geq|\langle\sum_{J=1}^{N-1}Pr_{Y}^{B}(J)\rangle|/2\label{eq:Spin UR}
\end{equation}
The hidden states $(\lambda_{Z},\lambda_{X},\lambda_{Y})$ (which
specify predetermined \emph{nonzero} values for each of these observables)
contradict (\ref{eq:Spin UR}). As with inequality (\ref{eq:epruncert})
this contradiction signifies an EPR-steering of the cat-system and
negates any mixture of (local) quantum ``dead'' and ``alive''
hidden states consistent with LR \cite{supp}. 

So far, there is no falsification that the cat-system can be described
as a mixture of ``dead'' or ``alive'' \emph{hidden variable }states.
Such a falsification is achieved if the full LHV model (\ref{eq:lhv})
can be violated for the cat-state. This is predicted possible using
Svetlichny inequalities \cite{svet-1}. 
\begin{figure}
\includegraphics[width=0.33\columnwidth]{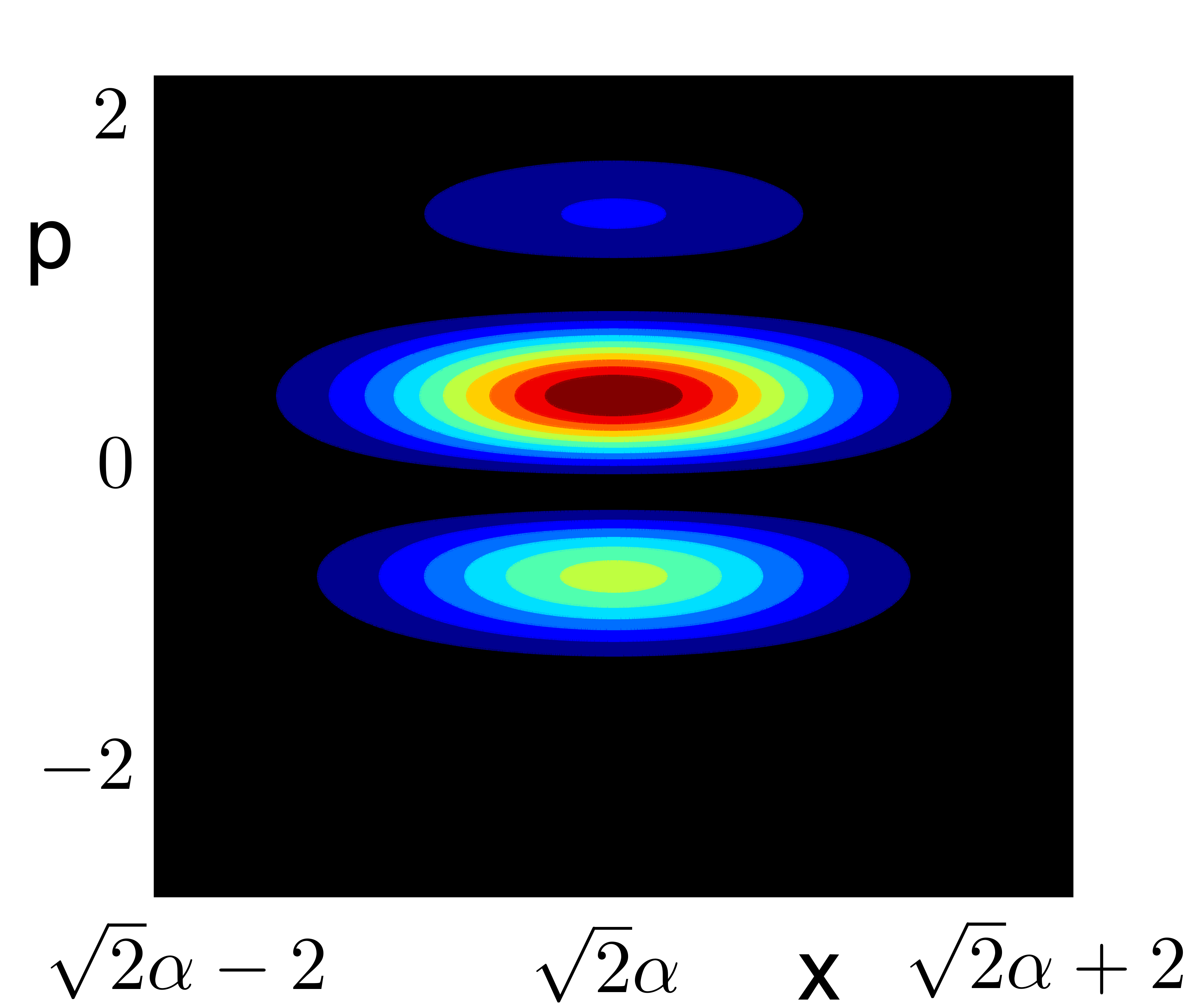}\includegraphics[width=0.33\columnwidth]{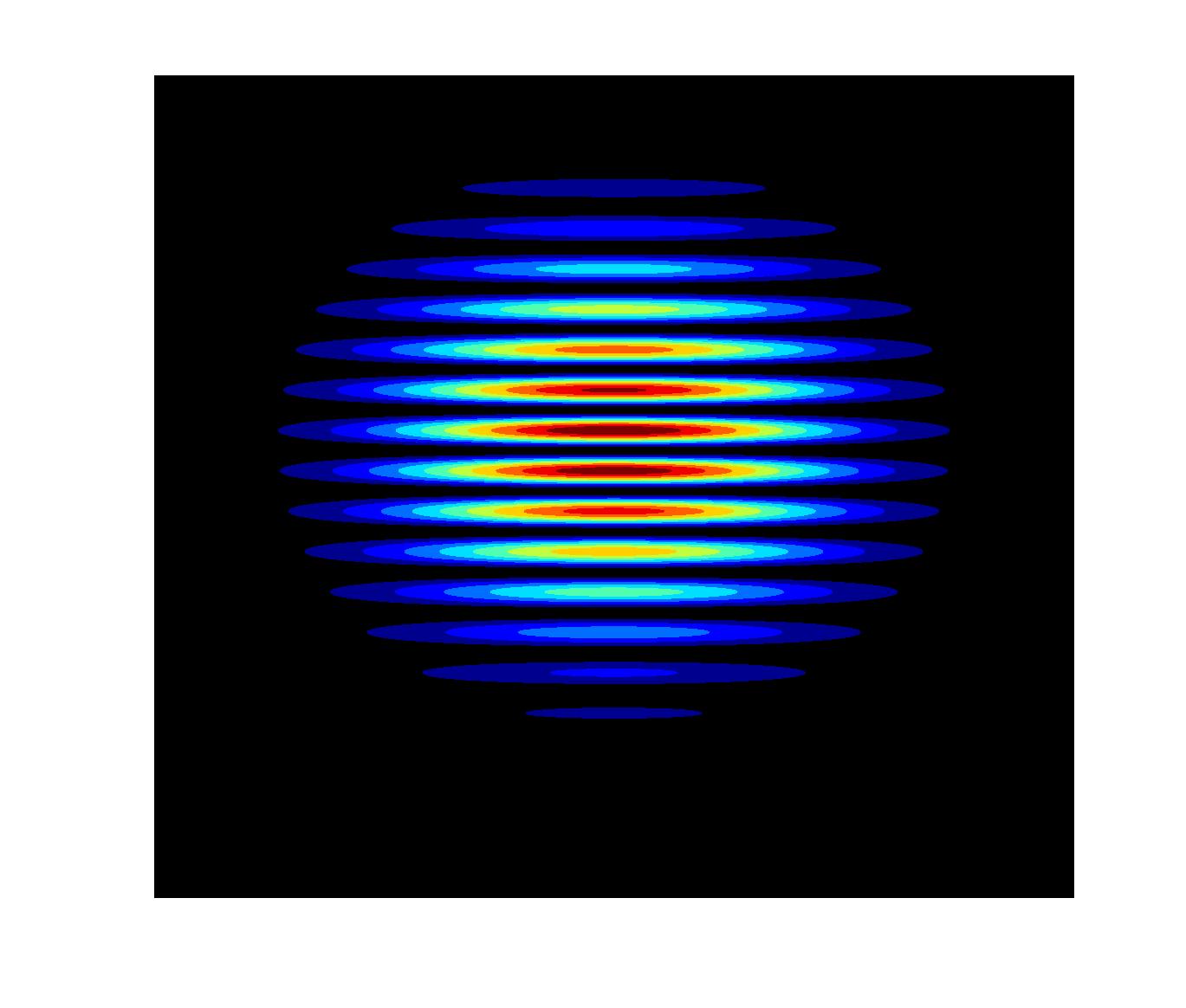}\includegraphics[width=0.33\columnwidth]{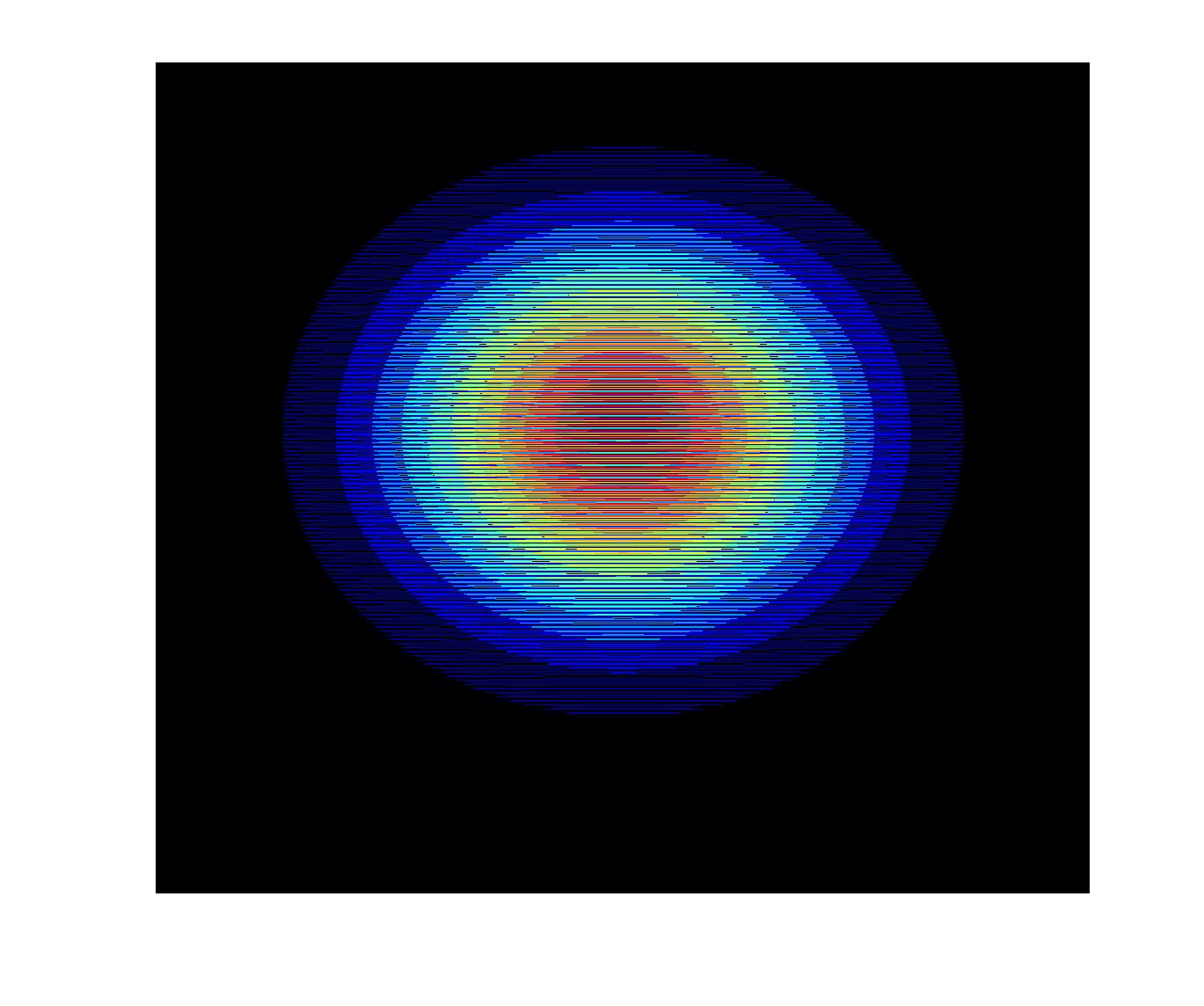}

\protect\caption{\emph{Quantum-classical transition:} Predictions of the element of
reality state $(1,1)$ for the cat-system of (2) with (from left)
$\alpha=2,10,100$. Plotted are contours for $P_{1}(x)P_{1}^{(X)}(p)$$ $
versus $x$ (horizontal) and $p$ (vertical). As $\alpha\rightarrow\infty$
the element of reality cannot be resolved as distinguishable from
the classical description $|\alpha\rangle$.}
\end{figure}

\textbf{\emph{Discussion}}:  The EPR steering signatures derived
in this paper falsify that the local cat-system can be in a ``dead''
or ``alive'' quantum state \emph{consistent} with the full separability
of the LHS model i.e. assuming no nonlocal effects between the ``cat''-
and spin systems.  Nonlocality has been verified in recent experiments
but such nonlocal effects are small, corresponding to predictions
of $\sim$ one spin unit. A relevant question is whether such small
nonlocal effects could explain the ``cat''-paradox in the context
of the state (1). Our results suggest this cannot be ruled out: If
we allow that there could be microscopic nonlocality, then the steering
signatures of the cat-states (2) and (3) vanish. To discuss this,
we \emph{quantify} the LR premise. We define that for\emph{ $\delta$-}scopic\emph{
LR $S\rightarrow C$,} it is assumed that Alice's measurement of the
spin does not affect the (value of measurement on the) cat-system
by an amount more than $\delta$. Hidden variables for the cat-system
 can then be defined with different amounts of indeterminacy $\delta$
in the prediction for measurements (due to different amounts of allowed
nonlocal change $\delta$).

\textbf{\emph{Result (2a):}} The EPR steering signatures are a negation
of a fully separable LHS model (\ref{eq:lhs}). We can determine a
value of $\delta$ such that if we allow nonlocality by an amount
greater than $\delta$, then the cat-system becomes indistinguishable
from the classical mixture.  We find $\delta$ (a measure of discrepancy
between the quantum and classical descriptions) is classifiable as
microscopic.

To explain, the EPR steering manifests in the cat-state (2) at large
$\alpha$ through very fine fringes in distributions for $P$. One
needs only relax the full locality condition to allow $\delta$-scopic
LR, where $\delta$ is a very small change in $P$, to nullify the
steering.  For the GHZ state, the EPR steering is lost when $\delta$
is a single spin unit. This ultra-sensitivity is consistent with
proven fundamental requirements for signifying macroscopic quantum
superpositions \cite{gisinsensi,peres}.

Moreover, we see from Figure 2 that as $\alpha\rightarrow\infty$
the signature requires an increasingly stricter form of LR i.e. the
value $\delta$ becomes smaller. This explains (similar to Refs.
\cite{peres}) the fragility to decoherence as the size $\alpha$
of the cat-state increases $-$ the ``Schrodinger cat''-like behaviour
is more difficult to observe because the elements of reality (which
give a predetermination of the results of measurement) are closer
to classically consistent values.

Hence it is the hidden variables/ states for the cat-system that specify
outcomes of measurement to a \emph{microscopic precision} that are
negated by the EPR-steering signatures. The hidden variable $\lambda_{Z}$
(that predetermines the outcome for the measurement $X$ or $\sigma_{Z}^{B}$
distinguishing the cat-system to be either ``alive'' or ``dead'')
can be defined with a macroscopic indeterminacy $\Delta$, in which
case we refer to it as a \emph{macroscopic hidden variable} $\tilde{\lambda}_{Z}$.
Because the ``dead'' and ``alive'' states are macroscopically
separated, the hidden variable $\tilde{\lambda}_{Z}$ can still predetermine
the cat-system to be ``dead'' or ``alive'' even for large $\Delta$,
though without full specification of the microscopic details of the
prediction. The variable $\tilde{\lambda}_{Z}$ requires the assumption
of $\Delta$-scopic LR, but this is implied by $\delta$-scopic LR
where $\delta<\Delta$. The Result (2a) indicates that for most typical
Schrodinger cat-type scenarios, the macroscopic hidden variable $\tilde{\lambda}_{Z}$
\emph{cannot be negated}. We can quantify with the following \cite{supp}.

\textbf{\emph{Result (2b):}} Suppose the uncertainty relation for
$X$ and $P$ is $(\Delta X)(\Delta P)\geq c$ where $c$ is a constant.
Suppose we assume $\Delta$-scopic LR and $\delta$-scopic LR to
deduce the hidden variables for measurement $X$ and $P$ respectively.
 If $\Delta\delta\geq c$, we cannot signify the cat-paradox by
negation of the LHV model based on $X$, $P$ measurements.$\square$
 This means that if we allow nonlocal effects of order $\gtrsim c$,
the signature of the cat-paradox is lost. As $\alpha\rightarrow\infty$,
the macroscopic element of reality $\tilde{\lambda}_{Z}$ predetermining\emph{
}the cat-system to be ``dead'' or ``alive'' can be defined with
$\Delta\rightarrow\infty$. But then $\delta$ needs to be increasingly
smaller.

\textbf{\emph{Conclusion:}} For typical scenarios modelling the macroscopic
entangled state (1), we cannot falsify the macroscopic element of
reality $\tilde{\lambda}_{Z}$ for the cat-system. The interpretation
of the cat-system being ``both dead and alive'' becomes debatable
in this context. This is clear because $\tilde{\lambda}_{Z}$ is precisely
the variable that \emph{predetermines} the outcome of the \emph{measurement}
distinguishing whether the cat-system is ``dead'' or ``alive''.
We can however falsify (by EPR steering inequalities) that the cat-system
is predetermined to be in a ``dead'' or ``alive'' \emph{local
hidden state,} where that hidden state has microscopic predictions
independent of measurements made on the spin system (as in the LHS
model (\ref{eq:lhs})). If the cat-system is a measuring-device pointer,
then an Ockham's Razor interpretation is illustrated by Figures 2b
and c. The pointer is positioned at one place on the dial \emph{or}
the other (as determined by the macroscopic element of reality $\tilde{\lambda}_{Z}$
that cannot be negated) but with its position / momentum microscopically
indeterminate due to microscopic nonlocality (illustrated by the fringes
that reveal the falsification of the LHS model). On the other hand,
we note from Result 2b that the quantification of the relevant nonlocal
effect is given by $c$ which more generally need not be considered
microscopic \cite{macrobell }.

The cat-states we describe can be realised for a mechanical oscillator
coupled to a two-level atom or optical system and for a microwave
field mode coupled to Rydberg atoms \cite{cats,ions}. Photonic states
 have been reported with Svetlichny-Bell violations \cite{catcqed,svetexp}.
The steering signatures are thus likely measurable by experiment.
\begin{acknowledgments}
I am grateful for support from the Australian Research Council Discovery
Program. I thank P Drummond for discussions and the 2016 Macroscopic
Entanglement Heraeus Seminar for the opportunity to present this work.\end{acknowledgments}

\end{document}